\documentclass[trackchanges, twocolumn]{aastex701}

\usepackage{amsmath}
\usepackage{amssymb}
\usepackage{cancel}

\usepackage{xcolor}

\newcommand{\uj}{\mathbf{u}_\mathrm{J}}
\begin{document}

\title{A microscopic analysis of shear acceleration in the relativistic regime} 

\author[orcid=0000-0002-4777-4842,sname='Aerdker']{Sophie Aerdker}
\affiliation{Ruhr-Universität Bochum, Fakultät für Physik und Astronomie, Universitätsstraße 150, 44780, Bochum, Germany}
\affiliation{Astroparticule et Cosmologie, CNRS, Université Paris-Cité, 75013 Paris, France}
\affiliation{Ruhr Astroparticle and Plasma Physics Center (RAPP Center), Bochum, Germany}
\email[show]{sophie.aerdker@rub.de}  

\author[orcid=0009-0009-3532-9686]{Xingwei Gong} 
\affiliation{Observatoire de Paris – PSL, Université PSL, 75014 Paris, France}
\email{}

\author[orcid=0000-0002-2395-7812,sname='Lemoine']{Martin Lemoine}
\affiliation{Astroparticule et Cosmologie, CNRS, Université Paris-Cité, 75013 Paris, France}
\email{}

\author[orcid=0000-0003-1334-2993]{Frank Rieger}
\affiliation{Max-Planck-Institut für Plasmaphysik (IPP), Boltzmannstraße 2, D-85748 Garching, Germany}
\email{}

\begin{abstract}

Relativistic shear flows, as encountered in jets of powerful astrophysical sources, offer promising sites of particle acceleration up to the highest energies. This stochastic acceleration process is usually described with Fokker-Planck transport equations, which assume small momentum gains per scattering. The present paper introduces a covariant microscopic description of relativistic shear acceleration, building on the formalism of generalized Fermi acceleration and formulating angular scattering in the locally comoving frame in terms of stochastic differential equations. This allows us to model relativistic shear profiles and to study the transition from gradual to non-gradual shear acceleration, where the particle mean free path becomes comparable to the shear length scale. 
Assuming an infinite shear layer, linear in the flow four-velocity, we observe that the shear rate peaks where the flow is mildly relativistic, whereas Lorentz aberration suppresses the transverse velocity of particles where the flow is relativistic. Consequently, particles that leave the central layer experience long waiting times with a heavy-tailed, power-law distribution. The momentum evolution becomes subdiffusive and the acceleration time can substantially exceed local estimates. We successfully reproduce the Monte Carlo energy distributions using a two-state continuous-time random walk with L\'evy-distributed waiting times. We also show how a momentum-dependent mean-free path can introduce breaks in the spectrum and extend our discussion to finite jet profiles. These effects should be accounted for in models of relativistic jets as sources of high-energy particles.
\end{abstract}


\section{Introduction}\label{sec:introd}
Fast shear flows are expected in a variety of astrophysical environments, including 
relativistic jets of gamma-ray bursts and active galactic nuclei (AGNs) \citep[e.g.,]
[for review]{Rieger2019}. Such shearing flows can give rise to shear acceleration of 
high-energy particles, a mechanism that has been studied in detail within relativistic
transport theory and stochastic acceleration frameworks, and that has been invoked to 
explain electron and cosmic-ray energization in astrophysical jets 
\citep[e.g.,][]{LiuRieger2017,WangReville2021,Tavecchio2021,Merten2021,Rieger2022,
WebbXu2023,HeSun2023,WangHuang2024,WangRieger2024,Tavecchio2025,WangHuang2025,RiegerDuffy2025,
WanSun2026,WangHe2026}. 
Recent kinetic and hybrid simulations of magnetized shear-driven turbulence further
show, in conjunction with MHD models, that shearing flows can self-consistently amplify 
magnetic fields, generate turbulent scattering environments, and energize non-thermal 
particles \citep{Wang+2023,2026ApJ..1009..247L,Liang+2026I, Liang+2026II}.

On the theoretical side, the general particle transport has been described by a mixed-frame, diffusive particle 
transport equation in which the scattering operator (or particle momentum) is evaluated 
in the comoving frame \citep{Webb1989,WebbJokipii1994,RiegerMannheim2002,WebbBarghouty2018}.
Although particle transport in such flows is generally space- and momentum-dependent, shear 
acceleration has been successfully modeled within a reduced, momentum-dependent Fokker-Planck 
framework \cite[e.g.,][]{RiegerDuffy2006,LiuRieger2017,RiegerDuffy2019,WangReville2021}, 
analogously to the classical second-order Fermi acceleration \citep{Fermi1949}. 
For a non-relativistic, incompressible shear flow $\uj(x)$, 
diffusion and drift in momentum space depend on the shear coefficient $\Gamma \propto 
(\partial u_{\rm J}/\partial x)^2$. Assuming homogeneous isotropic spatial diffusion and 
spatially uniform $\Gamma$, the corresponding Fokker-Planck equation reduces to a simple 
diffusion equation in momentum space \citep{Berezhko1982,EarlJokipii1988,RiegerDuffy2006}. 
For relativistic flow speeds, and in the absence of losses and escape of the system, 
shear acceleration leads to a quasi-stationary power-law momentum distribution $f(p) 
\propto p^{-s}$, where the spectral index  $s = 3 + \delta$ depends on the 
momentum-dependence of the scattering time $\tau(p) \propto p^\delta$ 
\citep{WebbBarghouty2018,RiegerDuffy2019,
RiegerDuffy2022}.
Interestingly, the acceleration time $t_\mathrm{acc}$ is inversely proportional to the 
mean-free path, in contrast to other Fermi processes like diffusive shock acceleration, 
making shear acceleration in astrophysical jets a promising source of ultra-high energy 
cosmic rays \citep[e.g.,][]{RiegerDuffy2022,WangReville2024}.

While earlier models focusing on gradual shear flows typically assumed a  flow 
profile such that $\Gamma(\boldsymbol{x}) = \text{const.}$, both the acceleration time 
and the power-law index of
the accelerated particle distribution generally depend on the specific form of the shear 
flow profile \citep{RiegerDuffy2022,RiegerDuffy2025}, requiring corresponding modifications 
for a Fokker-Planck treatment. Once the flow changes significantly over a particle mean 
free path (as in a velocity jump), a transition to non-gradual shear acceleration furthermore 
occurs \citep{JokipiiMorfill1990,Ostrowski1990,Ostrowski1998}.

In this paper, we present a complementary approach based on a generalized Fermi description 
\citep{Lemoine2019,Lemoine2021,Lemoine2025} to investigate shear acceleration in a 
relativistic setting. By tracking the particle momentum in the comoving frame at all times, 
this formalism provides a direct microscopic description of Fermi-type particle acceleration 
in structured or random velocity flows. In the comoving frame, the energizing forces acting 
on a particle are expressed in terms of the (laboratory frame) spatio-temporal gradients of 
the frame velocity, in particular its acceleration, compression and shear. Being covariant 
and non-perturbative by construction, this approach remains valid beyond the realm of the 
Fokker-Planck approximation, e.g., as expected at the highest energies where the mean free 
path becomes comparable or larger than the width of the shear layer, up to the asymptotic 
regime of non-gradual shear acceleration.
In the regime of application of a Fokker-Planck formulation, i.e., a small coherence time 
of the random force and small relative energy gain, this approach recovers the diffusion 
coefficient derived in the quasilinear approximation in earlier studies~\citep{Lemoine2025}.

For numerical applications, we formulate this generalized Fermi model in terms of stochastic 
differential equations, following the methods introduced in \citet{Merten-etal-2017,Aerdker-etal-2024,
MertenAerdker25, Luebke+2025}. 
The time evolution of the particle 4-momentum is then expressed in the comoving frame as a 
dynamical system combining the force term associated to the shear at the particle position 
with a stochastic module describing scattering.

This approach is well suited for numerical implementation by Monte Carlo methods, and it 
offers the additional flexibility to consider arbitrary flow profiles and scattering processes, 
e.g., anisotropic diffusion in space and space-dependent scattering times. Furthermore, as 
the displacements in space can be tracked as well, the evolution of the distribution in 
comoving momentum can be conveniently boosted back to the lab frame any time. 

In the first part of this paper, we consider an infinite shear profile in the $x$-dimension $\uj(x)$ 
and postpone to Sec. \ref{sec:discussion} the inclusion of boundaries and escape to apply our findings 
to a more realistic jet scenario. We also assume for simplicity three-dimensional isotropic diffusion, 
appropriate for large-amplitude turbulence ($\delta B /B \sim 1$). We consider a simple and generic 
linear shear profile in the jet four-velocity $u_{\rm J}(x) = u_0 + \tilde \sigma (x-x_0)$ with constant 
$\tilde\sigma$, but our results remain general, provided $\sigma(x)\equiv\gamma_{\rm J}^{-1}\partial_x u_{\rm J}$ 
depends on $x$; here and throughout, $\gamma_{\rm J}(x)$ represents the jet bulk Lorentz factor.

One of our main findings is that the time evolution of the particle spectrum departs from 
standard expectations in the relativistic regime. This follows from the observation that the 
shear acceleration rate is maximal in regions where $\vert u_{\rm J}\vert/c \lesssim O(1)$, as 
discussed in Sec.\ \ref{sec:model} below. Consequently, particles are efficiently accelerated in 
this region, but at a slower rate elsewhere\footnote{This observation bears some similarity to 
the recent study of \citet{2026ApJ..1009..247L}, although the shear layer is in that case 
sub-relativistic and localized in space because it is induced by a Kelvin-Helmholtz instability at a 
velocity discontinuity.}.  
Furthermore, particles spend extended periods of time outside the shear layer, to the point where 
the waiting time distribution to return to the shear layer takes a power-law form with a heavy tail. 
This effect arises because, in regions where $\gamma_{\rm J}>1$, a particle transverse velocity, 
important for shear acceleration, is limited by $\sim c/\gamma_{\rm J}$ due to Lorentz aberration. 
This ``trapping'' phenomenon implies that the overall momentum evolution becomes sub-diffusive in 
time, characterized by rapid momentum evolution inside the shear layer where $\vert u_{\rm J}\vert/c 
\lesssim O(1)$, and extended waiting times outside. We discuss in detail how this impacts the 
acceleration time and the spectral shape of the distribution function, and provide an analytical 
framework for modeling the spectral evolution. 

Our discussion is organized as follows: we detail the numerical model in Sec.~\ref{sec:model} 
and present our results for the linear shear profile in Sec.~\ref{sec:results}, then discuss them 
in Sec.~\ref{sec:discussion}, relaxing the assumption of a constant mean-free path and applying 
our modeling to a more realistic jet scenario.

\section{Generalized Fermi Model}\label{sec:model}

We consider a relativistic shearing flow with 4-velocity profile $\uj = 
u_\mathrm{J}(x)\boldsymbol{e_z}$, with $\boldsymbol{e_z}$ the unit vector along the $z-$direction. 
We adopt a comoving frame that moves with the plasma velocity, in line with the assumption that 
magnetic perturbations responsible for scattering are carried with the flow. Our assumption of 
isotropic diffusion in the comoving frame implicitly assumes that the mean background field does 
not control spatial transport. We thus disregard this component in what follows. We note, however,
that the following results can be generalized to the case of anisotropic -- i.e., parallel 
vs. perpendicular -- transport relative to the background field $\mathbf{\langle B\rangle}$, which we plan to address in future work. The shear tensor is defined 
as $\sigma_{ij}\equiv {u_{\rm J}}_{\beta,\,\gamma}{e^{\beta}}_i{e^{\gamma}}_j$, with greek letters 
running from 0 to 3, and latin ($i$, $j$) running over $x$, $y$ and $z$~\citep{Lemoine2025}. In the 
present case, the tetrad components ${e^{\beta}}_i$ take values ${e^{\beta}}_x\,=
\,(0,\,\boldsymbol{e_x})$, ${e^{\beta}}_y\,=\,(0,\,\boldsymbol{e_y})$ and 
${e^{\beta}}_z\,=\,(u_{\rm J}/c,\,\uj/v_{\rm J})$, with $v_{\rm J}\equiv u_{\rm J}/\gamma_{\rm J}$ 
the jet 3-velocity, and $\gamma_{\rm J} = \sqrt{1+u_{\rm J}^2}$. The time component 
${e^\beta}_{t}\,=\,(\gamma_{\rm J},\uj/c)$ then ensures that 
$\eta_{\beta\gamma}\,{e^\beta}_a {e^\gamma}_b\,=\,\eta_{ab}$, with $\eta_{\mu\nu}$ the Minkowski 
metric. Assuming that $\uj$ depends solely on $x$, the only non-zero component of the shear tensor 
is $\sigma_{zx}=\gamma_{\rm J}^{-1}\,\partial_x u_{\rm J}$.

We express the angular components of the particle momentum through its projections on each axis, 
$\mu_i' \equiv  p_i'/|\bf{p}'|$ in the comoving frame. The evolution of the momentum $p'$ and angular 
variables $\mu'_i$ are then given by \citep{Lemoine2025}
\begin{align}
    \label{eq:diffeq}
    \dot{p'} &= -p'\, \mu'_x \mu'_z \sigma_{zx}  \,,  \\
    \label{eq:diffeq1}
    \dot{\mu}_x' &= \mu_x^{'2} \mu'_z \sigma_{zx} + \dot{\xi}_x \;,  \\
    \label{eq:diffeq2}
    \dot{\mu}_y' &=  \mu_x' \mu_y' \mu_z' \sigma_{zx} +\dot{\xi}_y  \;,  \\
    \label{eq:diffeq4}
    \dot{\mu}_z' &=  -\mu_x' \left(1 - \mu_z'^2\right) \sigma_{zx} + \dot{\xi}_z,\,
\end{align}
where the dot denotes differentiation with respect to time $t'$ in the comoving frame. From 
this equation, one derives a diffusion coefficient $D_{p'p'} \simeq  
{p'}^2\langle{\mu'}_x^2{\mu'}_z^2\rangle \sigma_{zx}^2\tau'$, where $\tau'$ denotes 
the characteristic scattering time in the comoving frame; the angular average becomes $1/15$ for 
an isotropic population, so that the diffusion coefficient matches that obtained in quasilinear 
theory~\citep[e.g.,][]{Rieger2019}. Note that this diffusion coefficient is often written in terms of $\gamma_{\rm J}^4 
(\partial_x v_{\rm J}/c)^2$ rather than $\sigma_{zx}^2$, but $\partial_x v_{\rm J} = \gamma_{\rm J}^{-3} \partial_x u_{\rm J} = \gamma_{\rm J}^{-2}\sigma_{zx}$, and hence the two formulations are identical.

The above dynamical system contains driving terms proportional to the shear $\sigma_{zx}$ and 
stochastic processes $\dot{\xi}_i$ governing angular scattering, whose specific definitions are 
given further below. In the absence of scattering, this system expresses the conservation of the 
particle momentum and angular variables in the laboratory frame, describing ballistic motion. 
Correspondingly, the angular variables $\mu'_i$ evolve under the influence of $\sigma_{zx}$ 
because the Lorentz transformation that connects the two frames evolves with position $x$.
This dynamical system also displays a fixed point at $\mu'_z=\pm 1$, $\mu'_x =0$, describing particle 
motion along the jet axis at constant energy.

Regarding the implementation of scattering, we first note that $\dot{\xi}_x$, $\dot{\xi}_y$ 
and $\dot{\xi}_z$ are correlated, since ${\mu'_x}^2 + {\mu'_y}^2 + {\mu'_z}^2=1$ at all times 
(the angular variables contain only 2 degrees of freedom). Our procedure can be summarized as follows. First, 
$\mu'_x$ is treated as a pitch-angle cosine along a reference direction (here $\boldsymbol{e_x}$), 
and $\xi_x$ is correspondingly modeled as an Ornstein-Uhlenbeck process with scattering 
frequency $\nu'=1/\tau'$,
\begin{equation}
    d\xi_x\,=\, -2 \nu'\mu'_x \,\mathrm{d}t' 
    + \sqrt{2\nu'\left(1-{\mu'_x}^2\right)}\,\mathrm{d}W_{\mu'_x,t'}\,
    \label{eq:dxix}
\end{equation}
${\rm d}W_{\mu'_x,t'}$ denoting a Wiener process. We model the evolution of $\xi_y$ and $\xi_z$ 
using an artificial gyrophase $\phi'$, that evolves as a Brownian motion with scattering 
frequency $\nu'_{\phi'}$,
\begin{equation}
    \mathrm{d}\phi' =  \sqrt{2\nu'_{\phi'}} \mathrm{d}W_{\phi',t'},\,
    \label{eq:dphi}
\end{equation}
with, here as well, ${\rm d}W_{\phi', t'}$ denoting a Wiener process. The definitions
\begin{equation}
    \mu'_y = \sqrt{1-{\mu'_x}^2}\cos\phi',
\qquad
\mu'_z = \sqrt{1-{\mu'_x}^2}\sin\phi',
\end{equation}
also define the increments ${\rm d}\xi'_y$ and ${\rm d}\xi'_z$ that are obtained by differentiating these 
relations and applying Itô's lemma to Eqs.~(\ref{eq:dxix}) and (\ref{eq:dphi}). For the choice $\nu'_{\phi'}=2\nu'$ scattering leads to isotropic diffusion in space $\kappa_\parallel = \kappa_\perp = c / (3 \nu')$. 

\begin{figure}[htbp]
    \centering
        \includegraphics[width=0.9\linewidth]{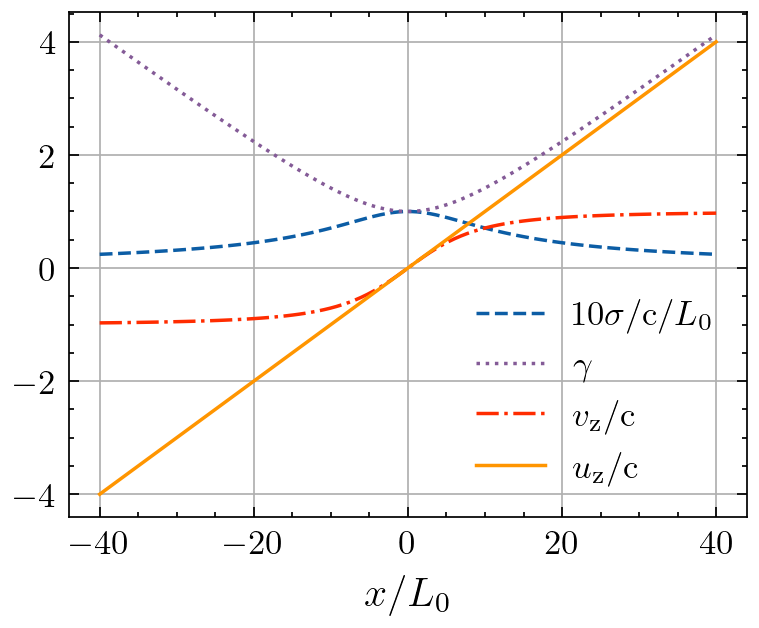}
    \caption{Profiles of $\uj(x), v_{\rm J}(x), \gamma_{\rm J}(x), \sigma(x) = \tilde{\sigma}/\gamma_{\rm J}$ as function
    of $x$, for constant $\tilde{\sigma} = 0.1\,c/L_0$ and  $u_0/c = 0$. }
    \label{fig:shearprofile}
\end{figure}

For the sake of clarity, we consider first an infinite linear shear profile in 4-velocity 
$\uj$, as illustrated in Fig.~\ref{fig:shearprofile}, i.e. 
\begin{align}
    \uj(x) &= \left(u_0 + \tilde{\sigma} x \right)\, \boldsymbol{e_z}\,,
\end{align}
where $\tilde{\sigma} = \partial_x u_{\rm J} = \mathrm{const.}$ and $u_0 
= u_{\mathrm J}(x = 0)$. Hereafter, we use the short-hand notation
\begin{equation}
    \sigma\,\equiv\, \sigma_{zx}\,=\,\frac{1}{\gamma_{\rm J}}\partial_x u_{\rm J}\,
    \label{eq:defsigma}
\end{equation}
that characterizes the acceleration rate, as discussed above. We also introduce a system-dependent 
unit of length $L_0$ so that, if $x$ is expressed in these units, $\tilde\sigma$ and $\sigma$ are 
expressed in $c/L_0$. Correspondingly, we express the scattering time in units of $\tau_0 = L_0/c$; 
$\sigma \tau'$ remains dimensionless.

The shear component $\sigma(x)$ depends on $x$, as illustrated in Fig.~\ref{fig:shearprofile}, and 
it reaches a maximum at $x_0  =  - u_0 / \tilde{\sigma}$. For the present shear profile, the 
characteristic length scale of the shear layer is best defined as $\ell_{\rm sh} \equiv c/\tilde\sigma$. 

A key observation is that $\sigma$ decreases in regions where the flow becomes relativistic, because 
$\sigma = \tilde \sigma/\gamma_{\rm J}$. This remains true for more general velocity profiles on 
dimensional grounds, since $\ell_{\rm sh}$ expresses the length scale over which $u_{\rm J}/c$ changes 
by an order unity relative to its value at the origin. Exponential profiles with $u_{\rm J}(x) \propto 
\exp\left(-x/\ell_{\rm sh}\right)$ are an exception, as they lead to $\sigma = u_{\rm J}/(\gamma_{\rm J}\ell_{\rm sh})$.

Particles will diffuse away from $x = x_0$ where the shear is strongest 
and acceleration is most efficient. We further distinguish the cases of a gradual shear acceleration 
when $\tau'\sigma < 1$ and $\tau'\sigma > 1$ towards non-gradual shear acceleration. The typical 
momentum gain per interaction is of order $O(\sigma \tau')$, thus, the 
small-gain assumption underlying a Fokker--Planck description becomes questionable when $\sigma\tau'$ 
exceeds unity.

\section{Results}\label{sec:results}

\subsection{Benchmarking on a classical Fokker-Planck model} 
\label{ssec:Fokker-Planck}

We first validate our models by comparing to the classical Fokker-Planck case. Assuming isotropic spatial 
diffusion, $\sigma\tau'\ll1$  and a spatially uniform value of the shear coefficient, the energy 
gain by shear acceleration can be described by a diffusion equation in momentum space. With the 
present choice of the linear flow profile $\mathbf{u}_\mathrm{J}(x)$, the shear $\sigma(x) = 1/
\gamma_\mathrm{J} \partial_x u_{\rm J}$ is dependent on $x$. To achieve a constant shear coefficient, 
in the simulation the particles' positions are set back to $x = 0$ after each integration step. 
In doing so, particles experience an approximately constant value of $\sigma$ around the initial value $x = 0$.  

Throughout this paper, our particle spectra $n_p$ in the laboratory frame, and $n_{p'}$ in the local comoving frame, are respectively defined as $n_p = p^2 \int_0^t{\rm d}\tau \int{\rm d}\Omega _p\,f(\boldsymbol{p},\tau)$ and $n_{p'} = {p'}^2\int_0^t{\rm d}\tau \int {\rm d}\Omega_{p'}\,f(\boldsymbol{p'},\tau)$ in terms of the distribution function $f$; ${\rm d}\Omega_p$ and ${\rm d}\Omega_{p'}$ are solid angle elements integrating over the angular distributions in momentum space. We plot our results at a given time $t$ in the laboratory frame, corresponding to different individual particle  proper times. Importantly, our particles are all injected at time $t=0$ and our spectra integrate over time; these spectra are then fully equivalent to instantaneous spectra calculated at time $t$ when injecting particles continuously and uniformly in time.

\begin{figure}
    \centering
     \includegraphics[width=1\linewidth]{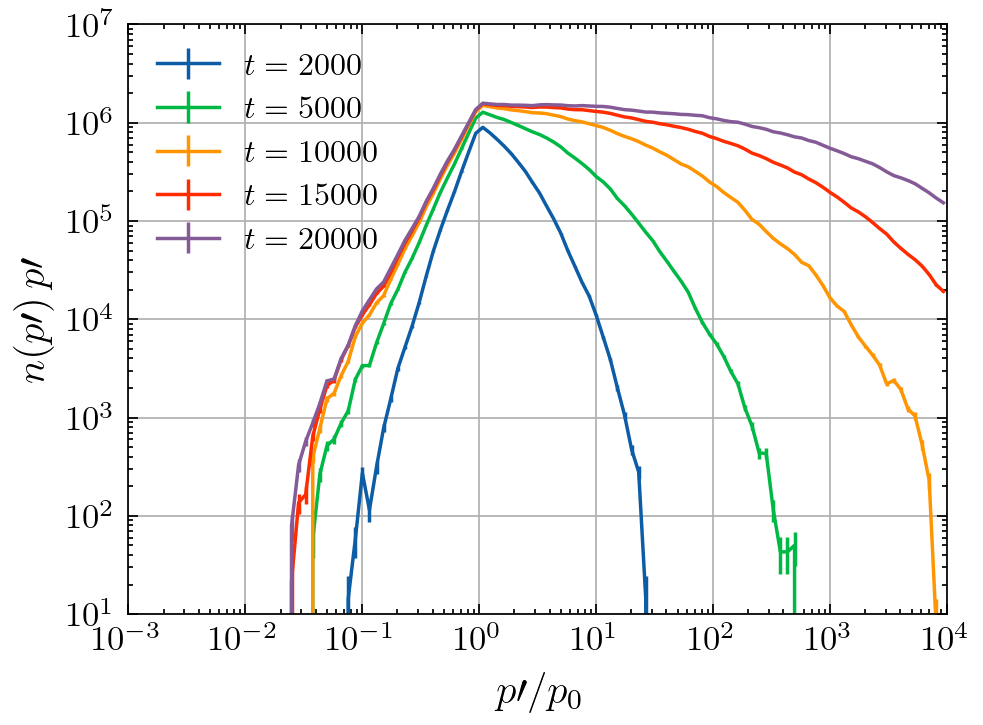}
    \caption{Particle spectrum $p'n(p')$, with $n(p')=4\pi {p'}^2 f(p')$ in terms of the distribution 
    function $f$, measured in the comoving frame for $\beta_0 = 0.2$, $\tilde{\sigma} = 0.1 \,c/L_0$ and 
    $\tau' = \tau_0$. Particles are reset every time step to $x = 0$ to ensure that they probe a 
    spatially uniform shear coefficient. The time asymptotic Fokker-Planck prediction $n(p') 
    \propto {p'}^{-1}$ is gradually recovered. The (laboratory) simulation time $t$ is expressed in 
    units of $\tau_0 = L_0 / c$.}
    \label{fig:Benchm_constScattering_reset}
\end{figure}

Figure \ref{fig:Benchm_constScattering_reset} shows the spectra in the comoving frame 
for $\beta_0 = 0.2$, $\tilde{\sigma} = 0.1\,c/L_0$ and $\tau'= L_0/c$, where $u_0 = 
\beta_0\,c/\sqrt{1-\beta_0^2}$. For a constant scattering time $\tau'$ and for a uniform shear 
the Fokker-Planck approach predicts a (time-integrated) power‑law spectrum that asymptotically 
approaches $n(p') \propto p'^{-1}$ when particles are continuously injected in the shear layer
\citep[e.g.,][]{RiegerDuffy2006}. 

The local (comoving) acceleration time scale derived in the microscopic Fokker-Planck approach 
is \citep[e.g.,][]{Rieger2019}
\begin{equation}
    \label{eq:acc_time_FP}
    t'_\mathrm{acc} = \frac{15}{(4 + \delta) \sigma^2 \tau'} \quad,
\end{equation}
where we have expressed the shear coefficient in terms of $\sigma = \tilde \sigma/\gamma_{\rm J}$, 
and $\delta$ denotes the momentum dependence of the scattering time: $\tau'\propto {p'}^\delta$; here $\delta =0$.
Evaluating the expression at $x=0$, with the corresponding values of $\beta_0$ and $\sigma$ yields 
a reference acceleration timescale of $t'_\mathrm{acc} = 390\,\tau_0$. 

\subsection{Gradual shear acceleration: $\tau'\sigma \ll 1$}
\begin{figure}[htbp]
    \centering
     \includegraphics[width=1\linewidth]{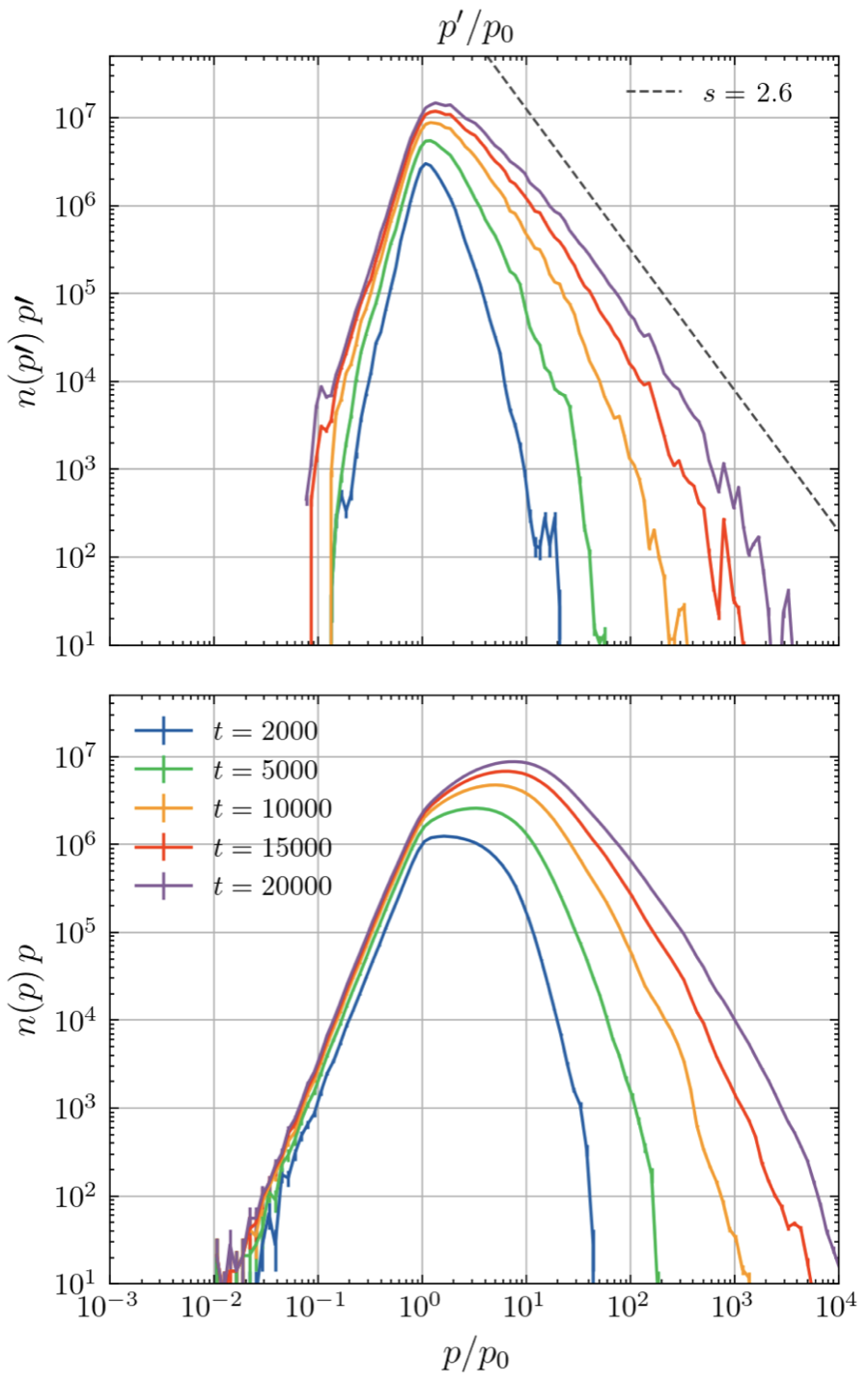}
    \caption{Particle spectra for the same parameters as in Fig.~\ref{fig:Benchm_constScattering_reset}, 
    namely $\beta_0 = 0.2$, $\tilde{\sigma} = 0.1 \,c/L_0$ and $\tau'=\tau_0$, albeit without restriction 
    on particle transport in the infinite linear shear profile. Particles are continuously injected at $x = 0$ 
    with $p' = p_0$. 
    Top: spectra in comoving frame momenta, bottom: spectra in lab frame momenta. The dashed line provides an 
    indicative power-law slope for $n(p') \propto p'^{-s}$ at $t = 2 \times 10^4\,\tau_0$. The hardening of the spectrum 
    in the lab frame at $p\sim 1-10\,p_0$ results from the Lorentz transformation from the comoving frame 
    to the lab frame.
    }
    \label{fig:Benchm_constScattering}
\end{figure}

In a next step we investigate the case of gradual shear acceleration in the infinite shear profile 
without constraining the spatial transport. The mean free path of the particles is chosen small 
compared to the size of the shearing region and independent of momentum ($\delta=0$) to reduce the 
parameter space. In line with earlier studies, we assume that particles are continuously injected 
at $x = 0$ with momentum $p_0$. 
Figure \ref{fig:Benchm_constScattering} now shows the time-integrated spectra $n(p')$ in the 
comoving and $n(p)$ in the lab frame for $\beta_0 = 0.2$, $\tilde{\sigma} = 0.1\,c/L_0$, and 
$\tau' = \tau_0$ as obtained from our generalized Fermi formulation. The spectra are integrated over 
space. 

In our Monte Carlo simulations, the particle position inside the shear profile is tracked as well, so 
that at any time the boost back to the lab frame can be performed. We do so by keeping track of the 
individual comoving time of each particle along its trajectory, and by reconstructing in the lab frame the 
particle spectra at a given lab frame time. The Lorentz transformation introduces a slight hardening 
in the lab frame spectrum around $p_0$, as particles are injected at $x = 0$ with $p' = p_0$ with random orientation.

The comparison between Figs.~\ref{fig:Benchm_constScattering} and \ref{fig:Benchm_constScattering_reset}, 
which adopt similar shear parameters, underlines the role of the spatial dependence of the shear profile.  
When particles are allowed to diffuse spatially and thus explore the spatial dependence of both $\uj$ and 
$\sigma$ (Fig.~\ref{fig:Benchm_constScattering}), the spectra are observed to be significantly steeper than 
those obtained for a spatially uniform shear acceleration rate (Fig.~\ref{fig:Benchm_constScattering_reset}). 
Furthermore, in the absence of escape the particle spectrum keeps evolving, without seemingly reaching steady 
state even on a restricted momentum interval, as for the spatially uniform shear case.  

We can estimate a characteristic mean acceleration time from the simulation using the variance in the 
momentum distribution over time $\langle \Delta {p'}^2\rangle/2\langle\Delta t\rangle$, as 
$t_\mathrm{acc} \simeq \frac{1}{4} p'^2 \left(\langle \Delta {p'}^2\rangle/2
\langle\Delta t\rangle\right)^{-1}$. At time $\Delta t = 10^4\,\tau_0$, we measure 
$\langle \Delta {p'}^2\rangle/2\langle\Delta t\rangle \approx 2\times10^{-4}\,p_0^2/\tau_0$ for 
$p' \in [1, 10]\,p_0$, implying $t_\mathrm{acc} \approx  1250\,\tau_0$, significantly longer than the 
reference value $t'_\mathrm{acc} = 390\,\tau_0$. Note, however, that the comoving time $t'$ differs 
for each particle and, in general, is shorter than the lab time, so that $t'_\mathrm{acc} < t_\mathrm{acc}$. 

Accounting for spatial dependencies leads to spectral softening and spatially varying acceleration 
rates, contrasting with the homogeneous case. In particular, particle acceleration is faster where 
$\sigma(\boldsymbol{x})$ peaks, with direct consequences for phenomenology, as we will discuss further 
in Sec.~\ref{sec:spatial}.

The quasi-linear general relativistic diffusive particle-transport framework formally incorporates 
spatial diffusion and a spatially dependent shear term \citep{Webb1989}. This provides a framework 
for modeling more realistic shear profiles, although at the price of increased complexity, by enlarging 
the dimensionality of the diffusive transport equation \citep[e.g.,][]{Webb1989,WebbJokipii1994,
RiegerMannheim2002,WebbBarghouty2018,Webb2019}. The latter two references obtain analytical steady state 
solutions to this full transport equation, including particle escape at a jet boundary, for a restricted 
class of scattering operators and flow profiles. In general, numerical methods are required to determine 
the particle distribution \citep{RiegerDuffy2022}. 

To preserve both the full spatial profiles of $\sigma$ and $\uj$ and the tractability of a 
one-dimensional (momentum) Fokker-Planck equation, one can average the full transport equation 
over the volume \citep{Webb2019,RiegerDuffy2019,RiegerDuffy2022}.
This yields a diffusion equation with an effective escape term $n/\tau_{\rm esc}$. In our case, this 
can be understood as an escape from the region where acceleration is most efficient, even though there 
are no physical escape boundaries applied.  In this limit, the spectral slope $n(p') \propto p'^{-s}$ 
becomes sensitive to the maximal value of the flow speed particles experience. 

To compare our results with this formulation, we extract the maximal flow speed from the Monte Carlo
simulations. At $t = 10^4 \,\tau_0$, the standard deviation of the particle distribution along the 
shearing direction $x$ is $\Delta x_{\rm rms} \approx 35\,L_0$. For this value, 
$\gamma_{\rm J}(\Delta x_{\rm rms}) \simeq 3.9$, corresponding to an outer velocity $\beta_{\rm J} 
\simeq 0.966$. The apparent reduction in spread is due to a diminished transverse (relative to jet 
propagation) velocity in regions where $\vert\uj\vert/c>1$: as particles scatter and isotropize their 
pitch angle in the comoving frame (ignoring shear forces for now), the mean transverse velocity in the 
lab frame is reduced by $1/\gamma_{\rm J}$ via Lorentz aberration. 

Equivalently, writing $\Delta t \simeq \gamma_{\rm J}\Delta t'$, and assuming isotropic diffusion in 
the comoving frame with diffusion coefficient $c^2\tau'/3$, one anticipates a squared perpendicular 
displacement $\Delta x_{\rm rms}^2 \sim c^2\tau'\Delta t /(3\gamma_{\rm J}\Delta x_{\rm rms})$, which 
reproduces the observed spread.
The effective outer velocity can be translated into a stationary power-law index for a 
volume-averaged Fokker-Planck description with effective escape from the acceleration region. 
Following \citet[][eq. 21]{RiegerDuffy2022} this is approximately $s_n = \frac{1}{2} - 
\left[ \frac{9}{4} + 116\, \langle \beta\rangle \left( \ln \frac{1+\beta_{\rm 02}}{1-\beta_{\rm 02}}
\right)^{-2}\right]^{1/2}$. Using the corresponding, weighted average of the velocity profile, 
$ \langle \beta\rangle \simeq 0.8$ and identifying $\beta_{\rm 02}$ with the effective outer velocity 
$\beta_J(\Delta x_{\rm rms})$ yields $s \simeq 2.3$, in reasonable agreement with the observed
spectral index, despite the lack of symmetry in the 3-velocity around $x=0$. The resulting spectrum 
is softer than in the constant-shear case modeled before. For concave-type shear profiles, 
\citet{Webb2019} derived an analytical solution based on the full relativistic transport equations, 
predicting a spectral index (their eq. 40) with $s = \frac{1}{2} - \sqrt{\frac{9}{4} + 20\,j_{\rm 0,1}^2 
\left[\ln\frac{(1+\beta_{\rm 02})}{(1-\beta_{\rm 02})} \right]^{-2}}$ with $j_{\rm 0,1} \simeq 2.4048$ (the first zero of the $J_0(z)$ Bessel function)
and $\beta_{\rm 02}:=(\beta_0-\beta_2)/(1-\beta_0 \beta_2)$. With $\beta_{\rm 02} = 
\beta_{\rm J}(\Delta x_{\rm rms})$, this yields $s \simeq 2.6$. While these results agree well with our 
findings in Fig.~\ref{fig:Benchm_constScattering}, they are predicted as an asymptotic value at a chosen 
time ($t=10^4\tau_0$), for the corresponding spread in position and effective flow speed. In particular, 
they cannot predict the spectral evolution that is observed in our simulations without further information 
on the particle spread.

Such spectral evolution takes place because particles retain a finite probability of returning to the 
central region where shear acceleration is maximally efficient, once they have escaped it. In the 
forthcoming section, we discuss this in further detail for the case $\sigma\tau'> 1$, and show that 
the particles' waiting time distribution outside the central region is furthermore power-law distributed 
with a heavy tail, leading to an overall sub-diffusive behavior in time. 
In actual astrophysical systems, one has to account for escape as well as a bounded shear profile, which 
both complicates the matter further. We investigate this issue further with a realistic jet profile in 
Sec.~\ref{sec:realprof}.

\subsection{Towards non-gradual shear acceleration: $\sigma \tau'\gtrsim 1$}
We now investigate a situation in which the mean free path becomes comparable or larger than the characteristic 
thickness of the shear layer. This scenario approaches the idealized case of a non-gradual shear, where particles 
move over a discontinuity in the velocity profile. 

\begin{figure}
    \centering
   \includegraphics[width=1\linewidth]{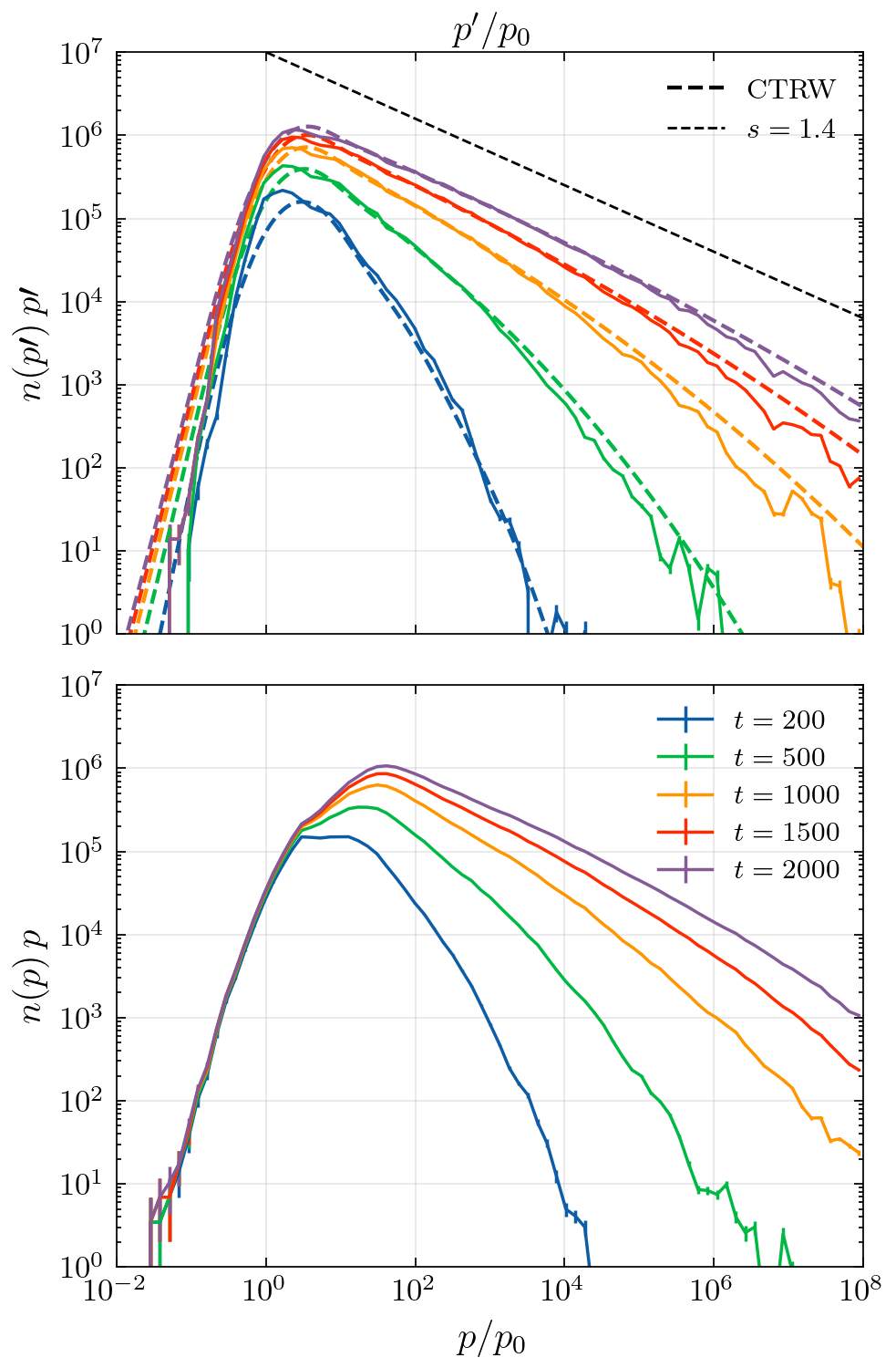}
    \caption{Particle spectra for $\beta_0 = 0.8$, $\tilde{\sigma} = 0.6 \,c/L_0$ and $\tau'=10\,\tau_0$. 
    Particles are continuously injected at $x = 0$ with $p' = p_0$. 
    Top: spectra in comoving frame momenta, bottom: spectra in lab frame momenta. The dashed line again 
    provides an indicative power-law slope for $n(p') \propto p'^{-s}$ at $t = 2 \times 10^3\,\tau_0$. 
    The colored dashed line shows the resulting time-integrated spectra from a 
    continuous-time random walk model (CTRW) for which we extracted the power-law of the waiting time distribution 
    from the simulation. 
    }
    \label{fig:spectra_coninj_subdiff}
\end{figure}

We expect shear acceleration to become more efficient with increasing mean-free path, as particles then experience 
a larger change in $u_{\rm J}$ over one scattering, and the typical momentum gain $\left< \Delta p\right> / p 
\sim  (\sigma \tau')^2$ exceeds unity once $\sigma\tau'  > 1$. At a given time, one consequently expects a 
significantly harder spectrum than in the previous case where $\sigma\tau'<1$. This is indeed confirmed by the 
time-integrated spectra shown in Fig.~\ref{fig:spectra_coninj_subdiff} for our adopted fiducial parameters 
$\beta_0 = 0.8$, $\tilde{\sigma} = 0.6 \,c/L_0$ and $\tau'=10\,\tau_0$.

\begin{figure}
    \centering
      \includegraphics[width=1\linewidth]{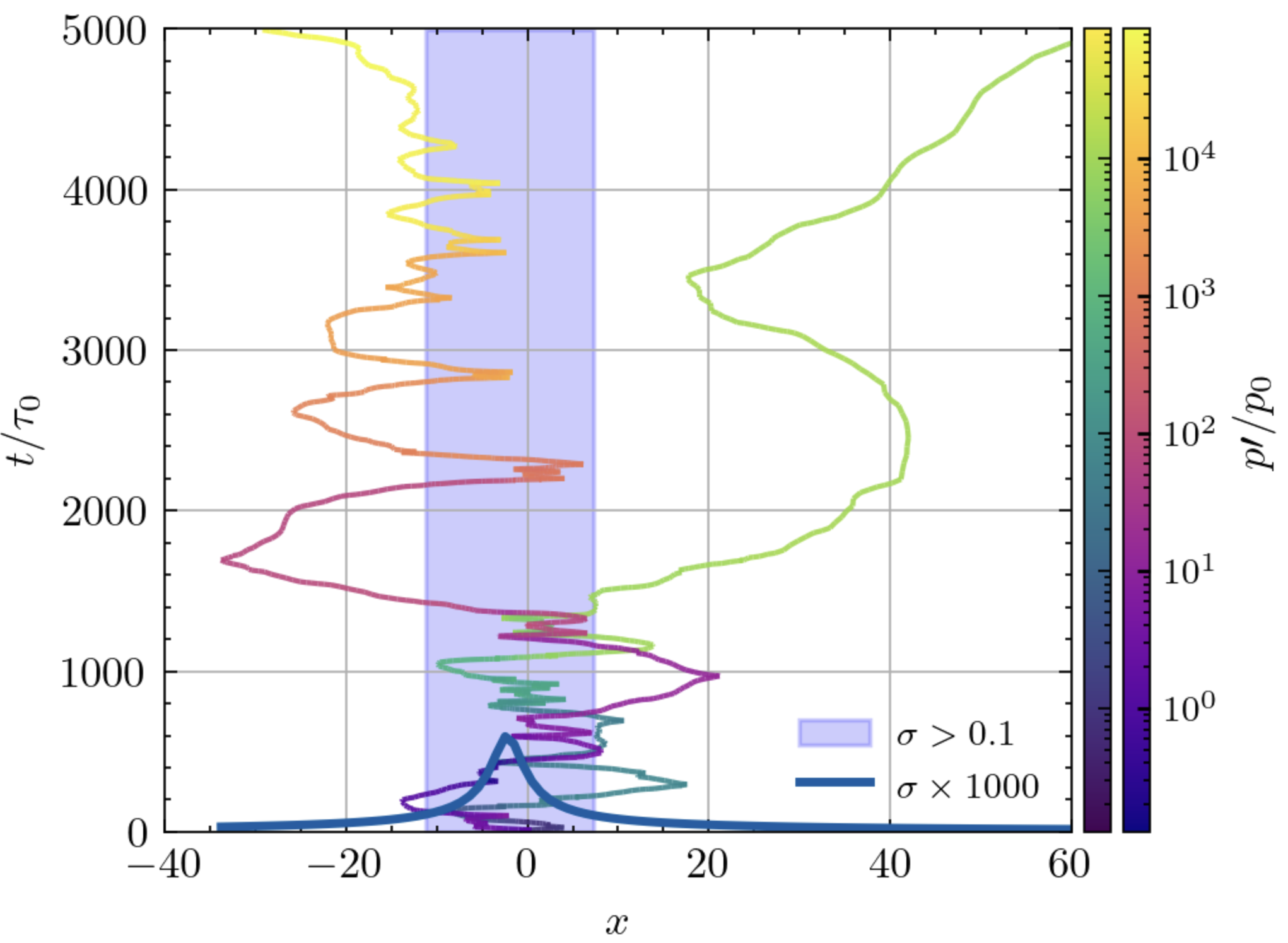}
    \caption{Example trajectories of two individual particles along the shearing direction $x$ and over time for our 
    strong shear case with $\beta_0 = 0.8$, $\tilde{\sigma} = 0.6\,c/L_0$ and $\tau' = 10\,\tau_0$.  The profile 
    $\sigma = \tilde{\sigma}/\gamma(x)$ is shown for comparison, the shaded area indicates the region of significant 
    shear, delimited by $\sigma > 0.1\,c/L_0$. The color tracks the comoving momentum $p'/p_0$. 
    Particles typically gain momentum when they are trapped in the shear layer while it remains quasi-constant once 
    they are outside. Note the difference in scale between the ordinates and the abscissa. }
    \label{fig:example-trajectories}
\end{figure}

\begin{figure}
    \centering
    \includegraphics[width=0.9\linewidth]{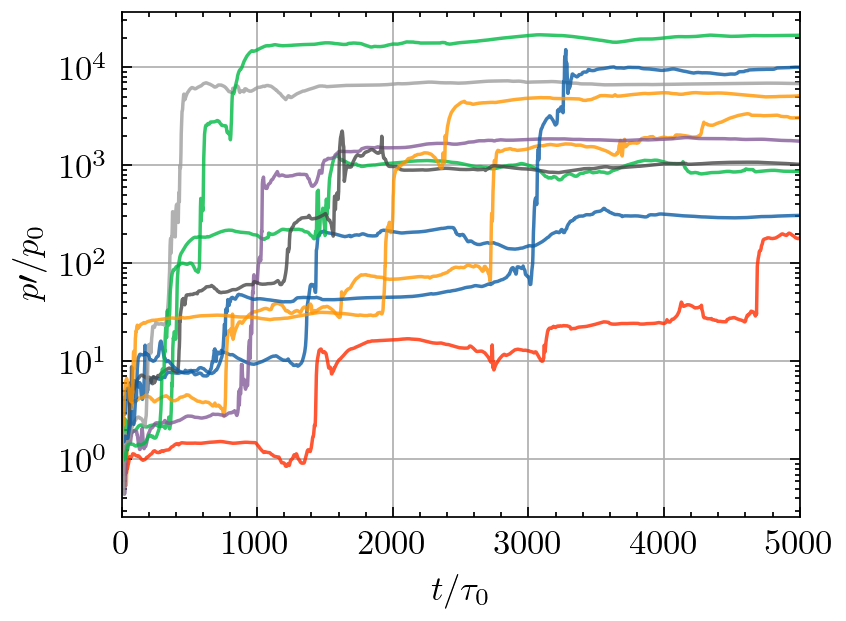}
    \caption{Example trajectories of the comoving momentum for ten particles as a function of time, for our strong 
    shear case with $\beta_0 = 0.8$, $\tilde{\sigma} = 0.6\,c/L_0$, $\tau' = 10\,\tau_0$. The momentum jumps in an 
    intermittent fashion each time the particle interacts with the central shear region, but stalls outside.}
    \label{fig:mom_trajectories}
\end{figure}

However, our simulations also reveal that acceleration is far slower than anticipated based on the above estimate. 
While Eq.~(\ref{eq:acc_time_FP}) predicts $t'_{\rm acc}\simeq 1\,\tau_0$ close to injection, the 
acceleration time measured from the simulation reads $t_\mathrm{acc} \approx 830 \,\tau_0$ (measured 
in the laboratory frame) at 
$t = 10^3 \,\tau_0$. 
We observe that particle acceleration is indeed fast in the central shear layer, but significantly slower once the 
particle travels in the outer regions. Figure~\ref{fig:example-trajectories} illustrates the time history of the 
comoving momentum along the particle trajectory, showing that energy gain predominantly takes place in the central 
shear layer. There, sudden jumps in momentum highlighted by a sudden change in color often correspond to a rapid 
change in direction along $x$. Outside this main shear layer, the comoving momentum does not evolve significantly 
and, furthermore, particles tend to move at sub-relativistic velocities along the shear direction in the laboratory
frame (compare the scales in ordinates vs that in abscissa). 
Correspondingly, the time histories $p'(t)$ shown in Fig.~\ref{fig:mom_trajectories} display periods of long
waiting times interspersed by sharp gains or losses. 
This overall behavior increases the average acceleration timescale with respect to that estimated in the central 
shear layer. 

The slow velocity, and corresponding long waiting times outside the shear layer, as seen in the lab frame,
derive from the previously described Lorentz aberration: as particles travel in regions of relativistic jet 
velocity ($\gamma_\mathrm{J}>1$), and scatter in the comoving frame, they move preferentially along the $z-$axis 
rather than along the shear direction, their transverse velocity being reduced by a factor $1/\gamma_\mathrm{J}$. 
Since the shear is expressed in terms of a velocity gradient in the laboratory frame, it is this transverse 
velocity rather than the comoving one that matters. 
For the case shown in Fig.~\ref{fig:example-trajectories}, 
one particle typically travels in regions where $x > 20$, corresponding to $\gamma_\mathrm{J} > 10$.

If one were to plot similar time histories in the previous weak (gradual) shear case, one would observe a more 
diffusive-like behavior of the momentum over time, even though particles eventually escape the central shear 
layer and then become subject to a weaker shear, as already discussed.

\begin{figure}[htbp]
    \centering
\includegraphics[width=1\linewidth]{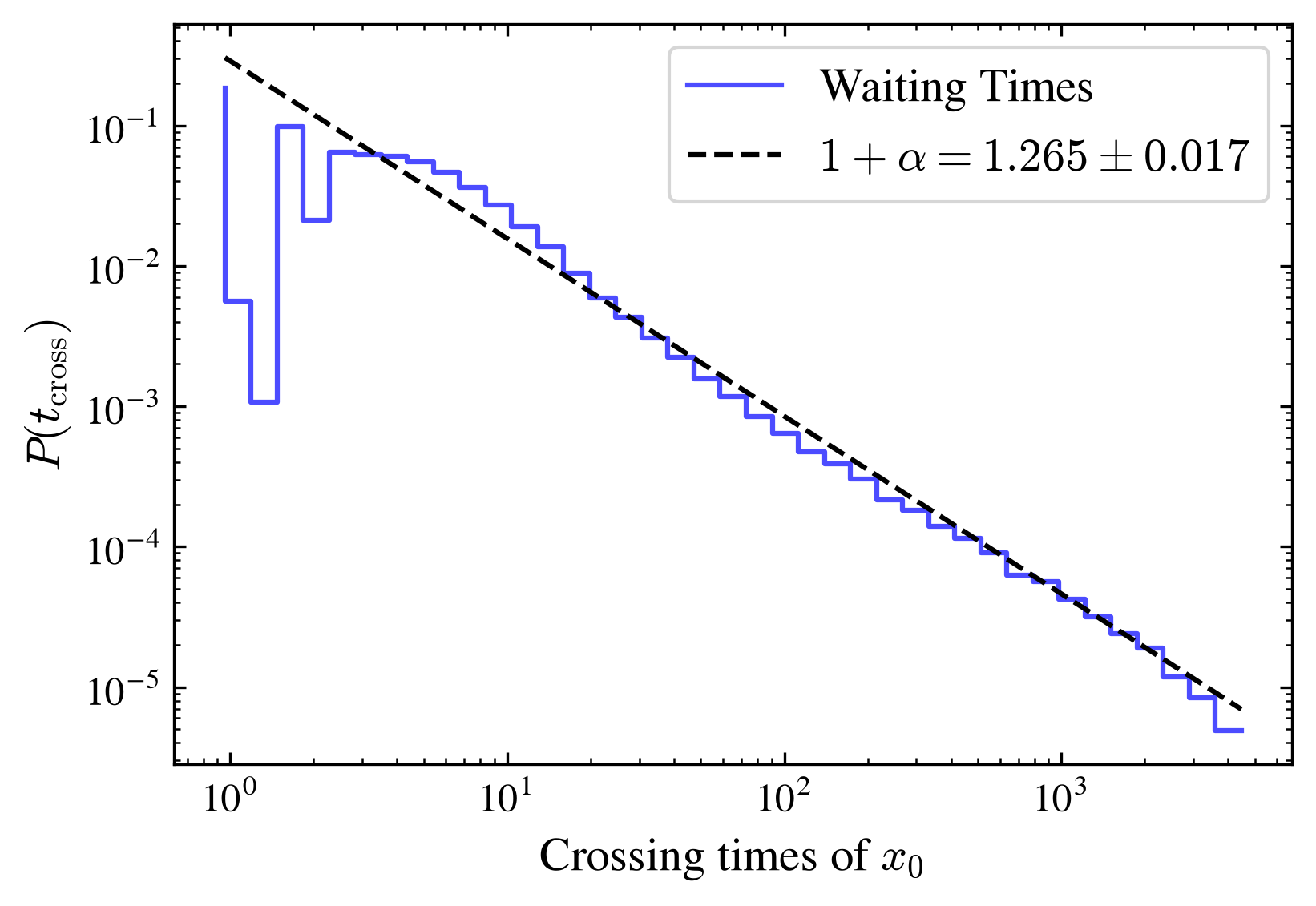}
     \caption{Waiting time distribution extracted from the simulation ($\beta_0 = 0.8, \tilde\sigma = 0.6\,c/L_0$), 
     between two crossings of $x_0$. 
     The power-law slope can be fitted to $-(1 + \alpha) \simeq -1.27$.
     }
    \label{fig:Waitingtimes}
\end{figure}

\begin{figure}
    \centering
        \includegraphics[width=0.9\linewidth]{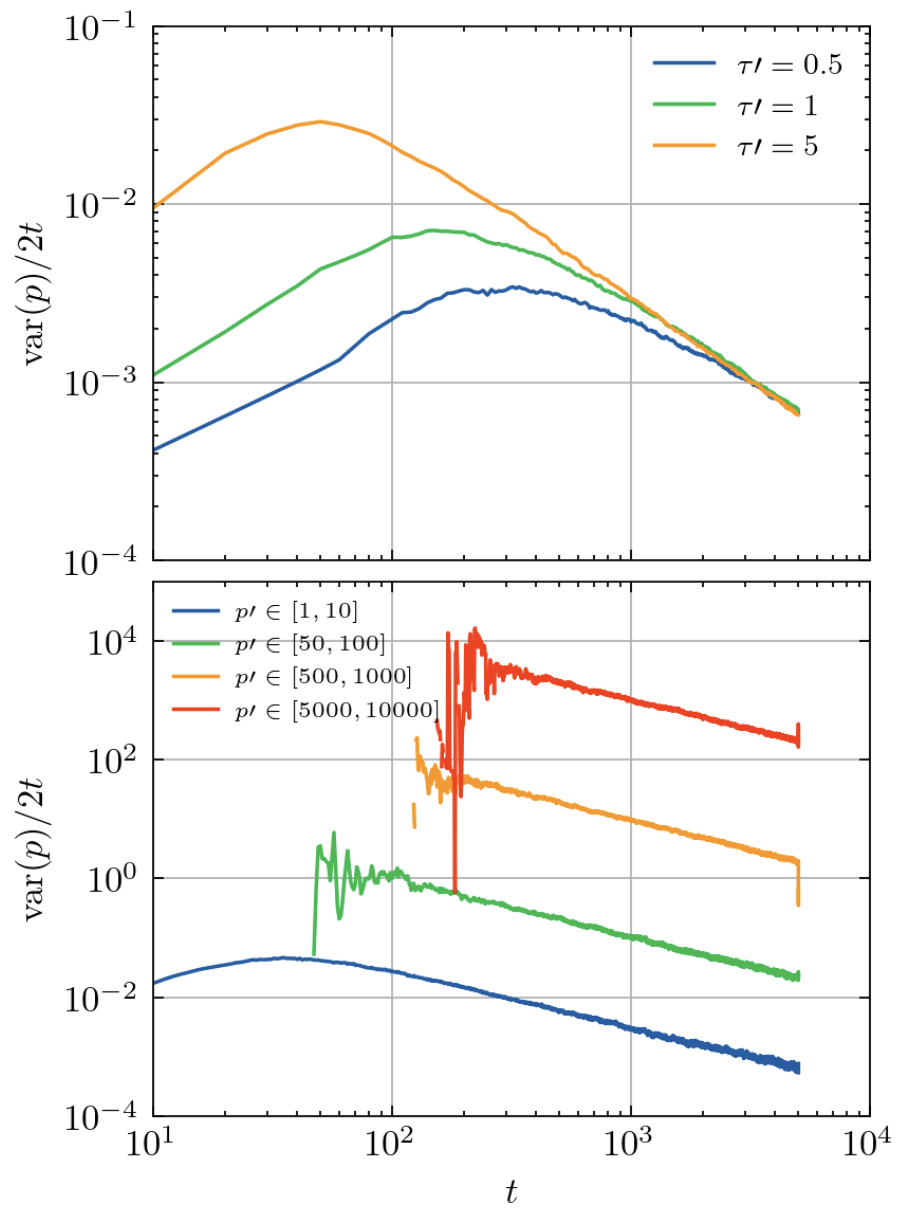}
    \caption{Top: Running momentum diffusion coefficient of the comoving momentum for $\beta_0 = 0.8$, $\tilde{\sigma} = 0.6\,c/L_0$ 
    and different scattering times $\tau'$; $p' \in [1, 10]\,p_0$. Bottom: Running momentum diffusion coefficient where $\beta_0 = 0.8$, 
    $\tilde{\sigma} = 0.6\,c/L_0$, $\tau' = 10\,\tau_0$ with respect to the laboratory time $t$ in units 
    of $\tau_0$. The process is subdiffusive with $\left<p'^2\right> \propto t^q$, $q < 1$. 
    }
    \label{fig:runn_diff}
\end{figure}

Random walks displaying long waiting times depart from a simple Brownian motion for which uncorrelated time steps take place at regular, short intervals. Our simulations reveal that this indeed the case here. In Fig.~\ref{fig:Waitingtimes}, we plot the probability distribution 
function of the waiting times $t_{\rm cross}$ experienced by the particles between two consequent crossings of $x_0$ where the shear is strongest. 
This distribution $p(t_{\rm cross})$ exhibits a tail $\propto t_{\rm cross}^{-1.27}$, which implies that the mean waiting time is not well defined, and that the random walk in momentum must be sub-diffusive.

We confirm this in Fig.~\ref{fig:runn_diff}, which shows the time evolution of running 
momentum diffusion coefficients $\langle \Delta p(t)^2\rangle/2t$. Whether plotted as a function of 
momentum (bottom panel) or scattering time (top panel), these plots systematically reveal sub-diffusive evolution 
of the momentum. 

A similar problem has been studied previously by \cite{LemoineMalkov2020} in connection with stochastic acceleration in turbulent plasmas. 
We thus borrow the tools developed there to interpret the observed particle distributions.
We build a simple model in which the time evolution of a particle momentum follows a two-state continuous-time random walk (CTRW). 
In state 1, particles travel in the central shear layer and undergo efficient acceleration for a fixed duration $\Delta t$,\footnote{Formally, 
$\Delta t$ is defined in the comoving frame, but we extend it to the laboratory frame, given that in our simulations the jet Lorentz factor 
is of order unity in the central shear layer.}, while in state 2, they travel for an extended time $t_{\rm cross}$ without undergoing acceleration. 
The probability distribution of $t_{\rm cross}$ is modeled by a L\'evy distribution $L_\alpha(\hat{t})$ where $L_\alpha(\hat{t}_{\rm cross}) 
\propto \hat{t}_{\rm cross}^{-1-\alpha}$ for large $\hat{t}_{\rm cross}$ and $\hat{t}_{\rm cross} \equiv t_{\rm cross} /\Delta t$ is the 
time between two acceleration events, normalized by $\Delta t$, the characteristic timescale of state 1. The parameter $\alpha$ is obtained 
by matching the waiting time distribution observed in our simulation, leading to $\alpha = 0.27$. The evolution of the particle momentum 
from momentum $p$ to $p'$ in state 1 is characterized by a Gaussian propagator for diffusion in momentum, $G_{\nu'}(p_1'|p_0',\Delta t)$. 
We use the standard form 
\begin{equation}
G_{\nu'}(p_1'|p_0', \Delta t) =
\frac{1}{\sqrt{4\pi\nu'\Delta t}}\exp\left[-\frac{\left(\ln(p_1'/p_0')-3\nu'\Delta t\right)^2}{4\nu'\Delta t}\right]
\end{equation}
that describes diffusive shear acceleration at constant (energy-independent) rate $\nu'$.

The probability that a particle undergoes $n$ waiting periods and $n$ acceleration episodes within time $t=\hat t\,\Delta t$ is given by 
\begin{equation}
    P_n(n,\hat{t},\alpha)
    = \int_{(\hat{t}-n-1) / (n+1) ^{1/\alpha}} ^ {(\hat{t} - n) / n^{1 / \alpha}}\mathrm{d} \hat\tau\, L_\alpha (\hat\tau).
    \label{eq:pn}
\end{equation}
The particle spectrum at a given simulation time can then be described as the sum of the propagators 
for $n$ acceleration events weighted by the probability that a particle undergoes $n$ acceleration-trapping cycles within the 
simulation time. For L\'evy-distributed trapping times this spectrum can be approximated by

\begin{equation}
\small
    p'^2 f(p',t) \simeq
    \sum_{n=1}^\infty P_n(n, \hat{t}, \alpha)
    G_{\nu'}(p'|p'_0,n\Delta t),
    \label{eq:levy_spectrum}
\end{equation}

accounting for contributions from one to infinitely many crossings. 

To fix parameters, we define the shear region (state 1) as $\Delta x = |x - x_0| < 3/\sigma_0$, which represents regions that have a
dominant contribution to the particle energy gain. The average time spent in this region is a few scattering times, $\Delta t\sim 50\tau_0$,
for the simulation with $\beta_0 = 0.8$, $\sigma_0 = 0.6$, and $\tau = 10$.
The initial position of particle injection should not affect the fitting of $\alpha$, because this averaged  acceleration time 
is dominated by particles with multiple crossings.

With $\nu'= 6.9\times 10^{-2}$, the CTRW reproduces the late-time comoving spectra and their spectral evolution in the simulation. This is  consistent with our definition of the shear region, since relating $\nu'$ to an effective shear coefficient $\sigma_\text{eff}$ via $\nu' = D_{p'p'}/p'^2 =  \tau\sigma_{\mathrm{eff}}^2/15$, gives $\sigma_\text{eff} \simeq 0.5\, \sigma_0$. 

The time-integrated spectra predicted by this model are shown in Fig. \ref{fig:spectra_coninj_subdiff} together with the spectra obtained in the generalized Fermi model.
Note that the width of the strong shear region is not uniquely defined and is rather a coarse-graining choice. The strong shear region should be sufficiently wide to contain most acceleration, while sufficiently narrow such that the acceleration inside can still be modeled with a single momentum propagator. This model can well explain the spectral evolution with a reasonable choice of $\Delta x$ between $3/\sigma_0$ and $5/\sigma_0$, with the resulting $\alpha$ slightly modified. 

\section{Discussion}
\label{sec:discussion}

The generalized Fermi Monte Carlo model developed above recovers the standard Fokker–Planck limit for spatially 
uniform shear and small momentum changes, while allowing us to follow particle transport through inhomogeneous 
relativistic shear profiles. 
So far the model is applied to a linear, infinite shear profile in the four-velocity, assuming constant, isotropic 
scattering in the comoving frame and the spectra are integrated over space. In the following, we relax some of these 
assumptions and briefly discuss the impact of momentum-dependent scattering times, particle distributions in space, 
as well as the role of particle escape in an exponential shear profile. 

\subsection{Spatial distribution}\label{sec:spatial}

\begin{figure}[htbp]
    \centering
 \includegraphics[width=1\linewidth]{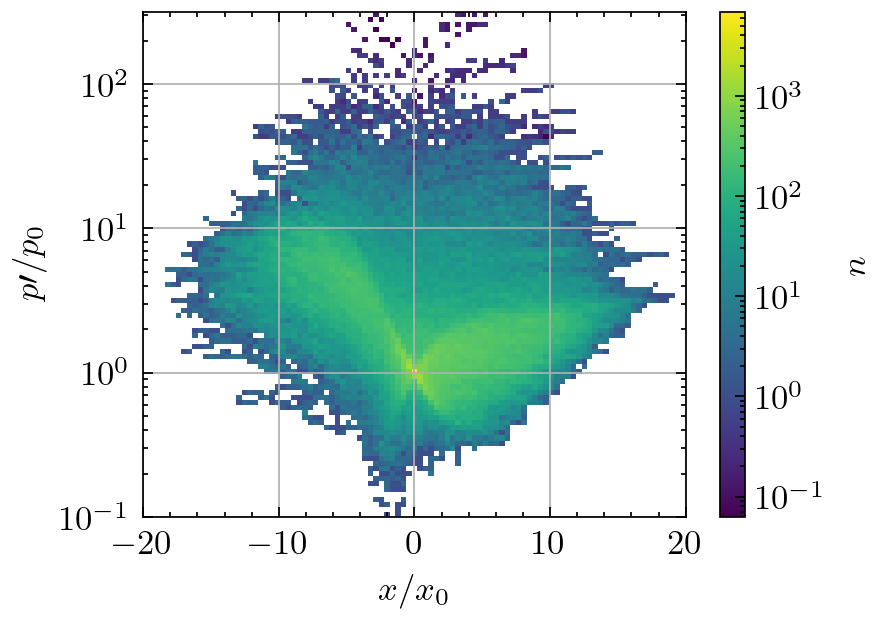}
    \caption{Particle number distribution along the shearing direction and momentum in the comoving frame $p'$ at 
    $t = 100\,\tau_0$ assuming continuous injection of particles at $x = 0$. The scattering time is constant with 
    $\tau' = 10\,\tau_0$, $\beta_0 = 0.8$, $\tilde{\sigma} = 0.6\,c/L_0$ (strong shear) and no escape 
    boundaries are applied. A number of $5000$ particles is injected and integrated in a Greens function approach 
    with simulation time step $\Delta t = 0.1\,\tau_0$. }
    \label{fig:dist_space}
\end{figure}

The generalized Fermi model also allows us to follow the spatial transport of
the particles. Figure~\ref{fig:dist_space} shows the particle distribution in
the $x$--$p'$ plane at $t=100\,\tau_0$ for the strong shear case
$\tilde{\sigma}\tau'\gtrsim1$, assuming a constant, momentum-independent
mean free path. The highest-energy particles are concentrated around
$x\simeq x_0$, where the shear $\sigma$ is largest and
acceleration is most efficient. These particles have crossed the shear layer
several times and therefore remain, at a fixed time, more localized in the
shearing direction than particles of lower momentum. The same is
seen for particles with $p'<p_0$, since significant momentum losses also occur during crossings of the shear layer.

The structure of the distribution depends on time. At very early times,
$t\sim\tau'$, particles that have moved appreciably in the $x$-direction
are precisely those that have crossed the shear layer and have therefore gained
or lost momentum. By contrast, particles moving mainly along the jet direction,
or along the $y$-direction, remain close to their injection momentum
$p'\simeq p_0$. Therefore the early $x$--$p'$ distribution shows a larger
variance in momentum at larger $|x-x_0|$, while particles near their initial
position remain around $p_0$. 
After several scattering times,
repeated crossings of the shear layer populate the high- and low-momentum
regions and the distribution evolves towards the one shown in
Fig.~\ref{fig:dist_space}. The substructure visible in that figure is
a remnant of this early-time evolution. At later times $ t > 100\,\tau_0$ the distribution extends
to both larger momenta and larger distances in $x$, but its overall shape
remains similar. The spatial spreading eventually slows once particles reach
regions of large $\gamma_{\rm J}$, where Lorentz aberration suppresses the
transverse velocity in the laboratory frame.

\subsection{Momentum-dependent scattering time}

So far, we have assumed for simplicity that the scattering time does not depend on
particle momentum. This limit may arise when non-resonant processes dominate the
scattering physics~\citep[e.g.][]{2024MNRAS.529L..13P}, or for perpendicular
transport once the particle gyroradius $r_{\rm g}$ exceeds the turbulence
coherence scale $\ell_{\rm c}$~\citep{2013MNRAS.430.1280P}. More generally,
one expects

$$
\tau'(p')=\tau'_0\left(p'/p_0\right)^\delta ,
$$
with $\delta\simeq0.3-0.5$ for resonant interactions in quasi-linear theory
\citep[e.g.][]{1989ApJ...336..243S} or for curvature scattering in
large-amplitude turbulence~\citep[e.g.][]{2023JPlPh..89e1701L,
2023MNRAS.525.4985K}. Bohm-like transport would correspond to
$\delta\simeq1$, while $\delta\simeq2$ may occur once
$r_{\rm g}>\ell_{\rm c}$~\citep[e.g.][]{2020MNRAS.498.5051R}. At a given value of $\sigma$,
a momentum-dependent $\tau'$ implies that particles can evolve from the gradual
regime, $\sigma\tau'\ll1$, into the non-gradual regime,
$\sigma\tau'\gtrsim1$, as they are accelerated.

\begin{figure}[htbp]
    \centering
    \includegraphics[width=1\linewidth]{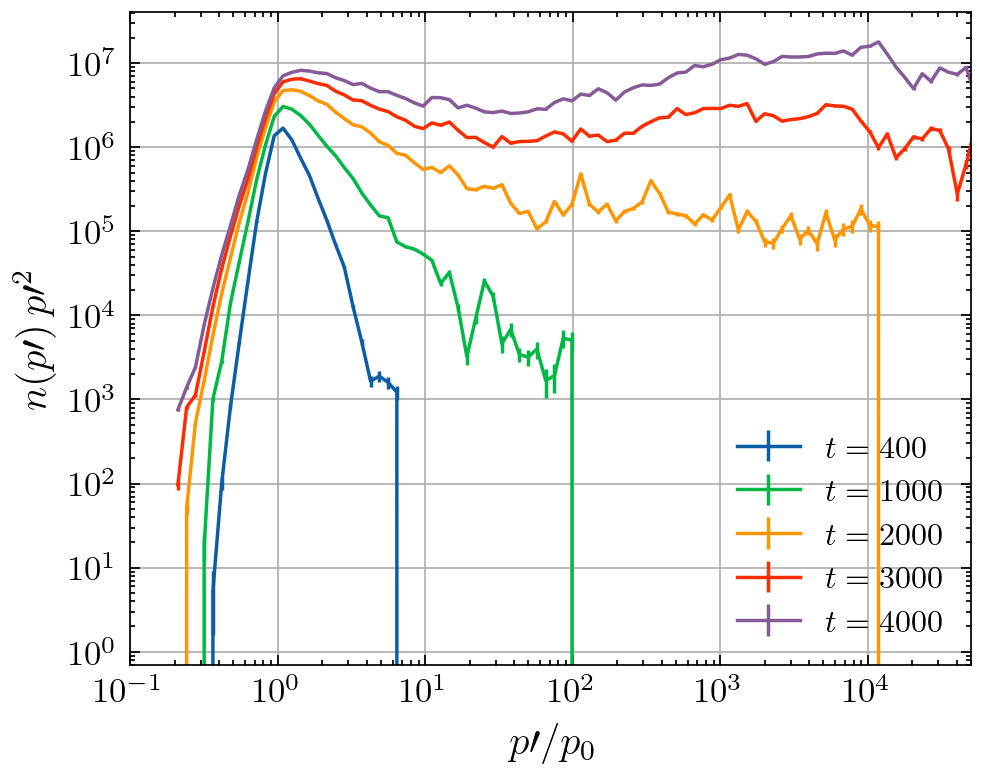}
    \caption{Spectrum with momentum-dependent scattering $\tau' = 1\,\tau_0 (p/p_0)$,  $\beta_0 = 0.2$, $\tilde{\sigma} 
    = 0.1\,c/L_0$ in the comoving frame.  For $p' > 10\,p_0$ the spectrum breaks into a harder spectra - similar 
    to the one previously observed in case of constant scattering in the limit of sub-diffusion.}
    \label{fig:GenFermi_d=1}
\end{figure}

Figure~\ref{fig:GenFermi_d=1} illustrates this effect for
$\tau'(p')=\tau_0(p'/p_0)$ and
$\tilde{\sigma}=0.1\,c/L_0$. The spectrum is shown as $p'^2n(p')$, so that a
horizontal curve would correspond to the homogeneous Fokker--Planck expectation
for $\delta=1$, namely $n(p')\propto p'^{-2}$, in the absence of escape and for a homogeneous velocity profile. For $p'\lesssim10p_0$, where
$\tau'\tilde{\sigma}\lesssim1$, the spectrum is broadly consistent with this
gradual-shear scaling. However, the spectrum keeps evolving in time and, analogous to our previous findings for a constant scattering time of $\tau = 1\,\tau_0$ the spectrum remains steeper even at late times $t = 4000\,\tau_0$.
Around $p'\simeq10p_0$, the condition
$\tau'\tilde{\sigma}\sim1$ is reached and the spectrum develops a break.

Above the break, the spectrum hardens instead of following the extrapolated
gradual-shear slope. This high-energy branch corresponds to
particles whose mean free path is large enough to sample a substantial velocity
contrast across the shear layer in one scattering time. In this sense, the
post-break spectrum resembles the hard, time-dependent spectra obtained in the
constant-$\tau'$ simulations with $\sigma\tau'\gtrsim1$. Its precise slope,
however, depends on the observation time,
the shear profile, and the statistics of returns to the acceleration region. 

Thus, a momentum-dependent mean free path can imprint a spectral feature at the
momentum for which $\sigma\tau'(p')\sim1$. Below this momentum, acceleration is
well described as gradual shear diffusion,  whereas above it the spectrum is shaped
by the crossing statistics of the shear layer in addition to the local
momentum-dependence of $\tau'$. 
Changes in the energy dependence of the spatial
diffusion coefficient, for example when $r_{\rm g}$ approaches or exceeds
$\ell_{\rm c}$, can therefore produce additional breaks or curvature in the
accelerated spectrum. This is directly relevant for ultra-high-energy cosmic rays
near the confinement limit, where $r_{\rm g}$ approaches the lateral size of the
shear layer or jet. In this regime, both the scattering law and the transition
from gradual to non-gradual shear acceleration may leave observable features in
the accelerated and escaping particle spectra. A detailed treatment including
spatially varying magnetic fields and losses is left for future work.

\subsection{Impact on shear acceleration in a jet profile}\label{sec:realprof}

\begin{figure}[htbp]
    \centering
    \includegraphics[width=0.9\linewidth]{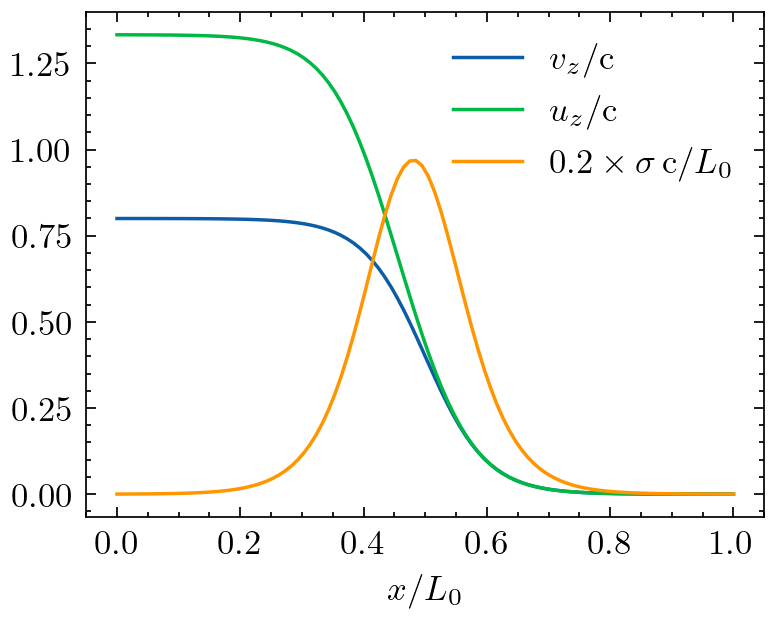}
    \includegraphics[width=1\linewidth]{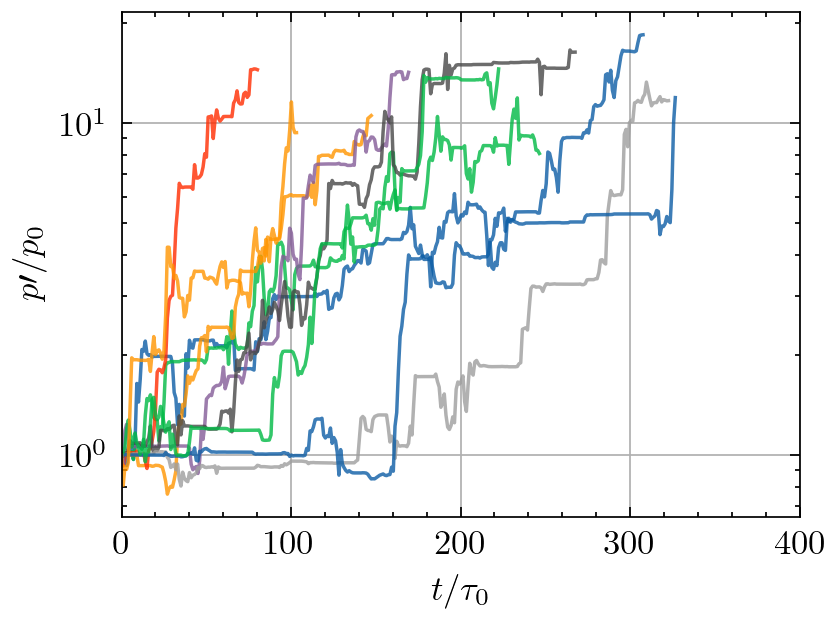}
    \caption{Top: Jet velocity profile adopted from Eq.~(\ref{eq:jet_profile}),
    inspired by \citet{Wang+2023}. The shear is maximal around $x_0$, where the
    flow velocity is still substantial. Bottom: Example trajectories of the
    comoving momentum for ten particles until escape. The parameters are
    $x_0=0.5\,L_0$, $w=0.05\,L_0$, and $\lambda=0.05\,L_0$, with a free
    escape boundary at $x_{\rm esc}=1\,L_0$.}
    \label{fig:Jet_profile-momentum}
\end{figure}

So far, we have studied the acceleration process in an infinite shear profile,
assuming a linear shear in the four-velocity and neglecting particle escape. We
now consider a first application to a finite jet-like profile, in order to assess
how the results change when the shear layer has a finite extent and particles can
escape. The scattering time is kept constant in this example. We adopt a
transverse velocity profile inspired by the AGN jet simulations of
\citet{Wang+2023},
\begin{align}
    v_z(x) =
    \frac{v_{z,0}}
    {1+\exp\left[(x-x_0)/w\right]} ,
    \label{eq:jet_profile}
\end{align}
with $x_0=0.5\,L_0$ and $w=0.05\,L_0$, as illustrated in
Fig.~\ref{fig:Jet_profile-momentum}. The velocity decreases from the jet
interior to the external medium over a layer of width $\sim w$, and the shear
is maximal around $x_0$. We impose a free-escape boundary at
$x_{\rm esc}=1\,L_0$. Particles moving toward the jet interior are reflected at
$x=0$, so that escape occurs only through the outer boundary.
Furthermore, particles are continuously injected with $p'=p_0$ uniformly
across the computational domain, $x\in[0,1]\,L_0$. This choice differs
from the previous case, where injection was localized near
the region of strongest shear.

For the parameters shown in Fig.~\ref{fig:Jet_profile-momentum}, the acceleration
remains intermittent. Particles gain most of their energy during crossings of the
shear layer, while intervals spent away from the layer correspond to little
momentum evolution. This behavior is analogous to that found in the infinite
linear profile, even though the finite jet geometry truncates the longest excursions and the jet is here mildly relativistic.
Consequently, the momentum evolution can still display subdiffusive features, but
the escape boundary prevents an indefinitely broad distribution of return times.
The strength of this effect depends on the ratio of the mean free path to the
shear-layer width and to the distance from the shear layer to the escape
boundary.

\begin{figure}
    \centering
    \includegraphics[width=1\linewidth]{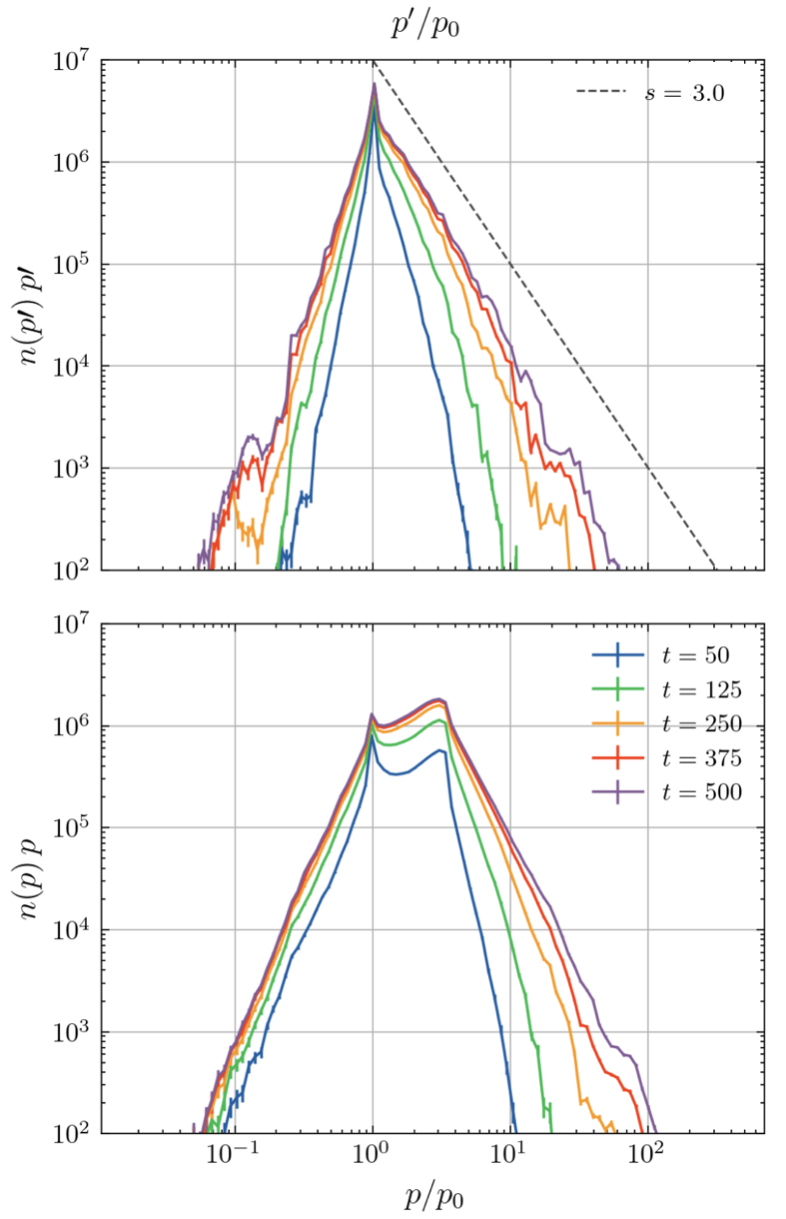}
    \caption{Particle spectra for the jet profile of Eq.~(\ref{eq:jet_profile})
    with $v_{z,0}=0.8c$ and $\tau'=0.02\,\tau_0$. Particles are continuously
    injected in the region $x\in[0,1]\,L_0$ with $p'=p_0$. Top:
    comoving-frame spectra; bottom: laboratory-frame spectra. The dashed line
    indicates the approximate power-law slope of the quasi-stationary spectrum at
    $t=500\,\tau_0$.}
    \label{fig:Jet_spectra}
\end{figure}

Figure~\ref{fig:Jet_spectra} shows the corresponding spectra for
$\tau'=0.02\,\tau_0$. In contrast to the unbounded profiles discussed above,
the finite escape boundary allows the system to approach a quasi-stationary state.
For the present parameters, the spectrum changes only weakly after
$t\simeq500\,\tau_0$, indicating that injection, shear acceleration, and escape
have reached an approximate balance. The resulting spectrum is softer than in the
unbounded case, because only particles that remain confined for several
shear-layer crossings populate the high-energy tail. In this sense, the spectral
index is controlled not only by the local shear-acceleration rate, but also by the
escape time and by the statistics of returns to the shear layer.
The dashed line in Fig.~\ref{fig:Jet_spectra} gives an indicative slope for this specific scenario of velocity profile, boundary condition and mean-free path. 

The peak at $p'=p_0$ reflects the injection prescription. Here particles are
injected throughout the full domain, $x\in[0,1]\,L_0$, rather than only near the
region of maximum shear. A fraction of the injected particles therefore starts
close to the escape boundary or in regions of weak shear and leaves the system
before significant acceleration occurs. When particles are injected near
$x=x_0$, nearly all of them will be accelerated and the peak at $p'=p_0$ (and the corresponding in the lab momentum spectrum) disappears.

\section{Summary}
In this paper, we have introduced a microscopic formulation of shear acceleration that is amenable to Monte Carlo simulations, and that can handle arbitrary flow profiles and differing scattering conditions (e.g., isotropic or anisotropic). The formulation tracks particles in a
mixed phase space, as in the relativistic transport formulation of
\citet{Webb1989}, and builds on the generalized Fermi framework of
\citet{Lemoine2025} by writing the dynamics as stochastic differential equations
in the local comoving frame.

In this first study we have focused on a linear idealized shear profile and isotropic diffusion with energy-independent scattering time to investigate the transition from gradual to non-gradual shear acceleration where $\sigma\tau'\gtrsim 1$ (with $\sigma$ the shear coefficient, $\tau'$ the comoving scattering time). We verified that the model recovers the standard
Fokker--Planck predictions for spatially uniform shear and small momentum changes
\citep[e.g.][]{Rieger2019}. Once spatial transport through the shear profile is
included, however, the ensemble evolution differs from that expected from a local
acceleration time alone. We have examined shear acceleration in a linear shear profile in flow $4-$velocity at different levels of shear strength, as quantified by $\sigma\tau'$.

We have observed that particle acceleration is most efficient in the region where the jet $4-$velocity is mildly relativistic, $\vert\uj\vert/c\sim O(1)$. This occurs as a combination of two effects. First, the shear acceleration rate scales as $\sigma^2 = \left(\gamma_{\rm J}^{-1}\partial_x u_{\rm J}\right)^2 $, which reaches a maximum in that region. Second, as particles isotropize their momenta through scattering in the comoving frame of the flow, their mean transverse velocity in the laboratory frame is reduced by $1/\gamma_{\rm J}$ because of Lorentz aberration. This implies that excursions in regions where the jet is relativistic take longer than naively anticipated, reducing the overall acceleration rate. For our idealized infinite shear profile, we observe that 
these 
``waiting times'' spent outside the region where acceleration is fastest are in fact distributed as a power-law with heavy tail, making the overall acceleration process subdiffusive in time. 
This effect becomes more pronounced as the shear strength $\sigma\tau'$ increases, or as the particle scattering mean free path increases. The acceleration process can thus be represented as a succession of energizing episodes in a central shear layer (where the jet velocity is mildly relativistic), interspersed by long excursions in regions of relativistic jet velocity, low acceleration rate. 
As we have shown, the time-integrated spectra can be reproduced by a model of stochastic acceleration using a continuous time random walk framework that incorporates the waiting time distribution. 
Thus, even when the local momentum diffusion coefficient is well defined
inside the shear layer, the ensemble-averaged acceleration rate can be controlled
by the return-time statistics of particles that have diffused into weak-shear
regions.

The waiting-time distribution is expected to depend
on the assumed velocity profile and boundary conditions. In a
finite jet, escape or a saturation of the velocity profile will truncate the
longest excursions and can drive the particle distribution toward a stationary
state. 
We have applied our method to the case of a more realistic jet profile, including net particle escape at a boundary, to test the robustness of our predictions. While the overall spectrum depends on the adopted geometry, as observed in previous studies, we still observe the emergence of sub-diffusive behavior once particles start to probe the inner relativistic regions. 

The transition to non-gradual shear acceleration is also sensitive to the
scattering law. If $\tau'$ increases with momentum, the parameter
$\sigma\tau'$ grows during acceleration, so that a particle population can move
from the gradual regime to an intermittent, non-gradual regime. This introduces an effective break in the particle spectrum at the momentum where $\sigma\tau'\sim 1$, with a transition from softer spectrum below to harder spectrum above. Such features need not be caused by
radiative cooling or escape alone, but may reflect a change in the character of
the acceleration process itself. This finding may have direct implications for the escaping spectrum of high-energy particles, and very-high energy cosmic rays in particular. We plan to address this issue in a forthcoming publication, using shear velocity profiles extracted from large-scale numerical simulations of relativistic jets.

The present calculations prescribe a constant or energy-dependent isotropic scattering time whereas in a realistic jet, it would likely vary in space.
This can affect the residence time in the accelerating region and the waiting time distribution, potentially enhancing particle trapping in inner regions if the transverse diffusion coefficient is suppressed there by, e.g., reduced turbulence amplitude and/or stronger field.
Radiative losses, especially for
electrons, will further compete with intermittent shear acceleration, while for ions the
dominant limitations are likely confinement and escape. 
Future applications should therefore combine the present approach with velocity, turbulence, and magnetic-field profiles extracted from
large-scale MHD or kinetic simulations of relativistic jets as briefly explored in sec.~\ref{sec:discussion}. 

\begin{acknowledgments}
The work of M.L. is supported by the French Agence Nationale de la Recherche, ANR, project ANR-25-CE31-3279.
F.M.R. acknowledges the ORIGINS Excellence Cluster, funded by the DFG under Germany’s Excellence 
Strategy–EXC-2094–390783311. S.A. acknowledges support by the Research Academy Ruhr and the RAPP Center, Bochum. 
\end{acknowledgments}

\bibliography{references}{}
\bibliographystyle{aasjournalv7}

\end{document}